\RequirePackage{fix-cm}
\RequirePackage{amsmath}
\documentclass[smallextended]{svjour3}       
\smartqed  

\usepackage{graphicx}
\usepackage{natbib}
\usepackage[utf8]{inputenc}
\usepackage{subcaption}
\usepackage[mathscr]{euscript}
\usepackage{amssymb}
\usepackage{makecell}
\usepackage{multirow}
\usepackage[export]{adjustbox}
\usepackage{color}

\usepackage{tikz}
\usetikzlibrary{arrows,positioning,shapes.geometric,matrix,graphs,chains,fit,calc}
\definecolor{myblue}{RGB}{80,80,160}
\definecolor{mygreen}{RGB}{80,160,80}

\newcounter{example1}[section]
\newenvironment{example1}[1][]{\refstepcounter{example}\par\medskip
   \noindent \textbf{RQ~\theexample. #1} \rmfamily}{\medskip}

\tikzset{ 
	mylabel/.style={draw=none,fill=none},
    table0/.style={
        matrix of nodes,
        row sep=-\pgflinewidth,
        column sep=-\pgflinewidth,
        nodes={
            rectangle,
            draw=black,
            align=center
        },
        minimum height=2em,
        text depth=0.5ex,
        text height=2ex,
        nodes in empty cells,
        row 1/.style={
            nodes={
                fill=mygreen,
                text=white,
                font=\bfseries
            }
        },
        column 1/.style={
            nodes={text width=3em,
            fill=myblue,
                text=white,
                font=\bfseries}
        }
    }
}

\tikzset{ 
    table/.style={
        matrix of nodes,
        row sep=-\pgflinewidth,
        column sep=-\pgflinewidth,
        nodes={
            rectangle,
            draw=black,
            align=center
        },
        minimum height=1.5em,
        text depth=0.5ex,
        text height=2ex,
        nodes in empty cells,
        row 1/.style={
            nodes={
                fill=black,
                text=white,
                font=\bfseries
            }
        },
        column 1/.style={
            nodes={text width=2.5em,
            fill=black,
                text=white,
                font=\bfseries}
        }
    }
}
  
\tikzset{ 
    table2/.style={
        matrix of nodes,
        row sep=-\pgflinewidth,
        column sep=-\pgflinewidth,
        nodes={
            rectangle,
            draw=black,
            align=center
        },
        minimum height=1.5em,
        text depth=0.5ex,
        text height=2ex,
        nodes in empty cells,
        row 1/.style={
            nodes={
                fill=black,
                text=white,
                font=\bfseries
            }
        },
        column 1/.style={
            nodes={text width=2.5em}
        }        
    }
}

\tikzset{ 
    table3/.style={
        matrix of nodes,
        row sep=-\pgflinewidth,
        column sep=-\pgflinewidth,
        nodes={
            rectangle,
            draw=black,
            align=center
        },
        minimum height=1.5em,
        text depth=0.5ex,
        text height=2ex,
        nodes in empty cells,
        row 1/.style={
            nodes={               
                text=red,
                font=\bfseries
            }
        },
        column 1/.style={
            nodes={text width=2.5em,
            fill=black,
                text=white,
                font=\bfseries}
        }
    }
}

\tikzset{ 
    table4/.style={
        matrix of nodes,
        row sep=-\pgflinewidth,
        column sep=-\pgflinewidth,
        nodes={
            rectangle,
            draw=black,
            align=center
        },
        minimum height=1.5em,
        text depth=0.5ex,
        text height=2ex,
        nodes in empty cells,
        row 1/.style={
            nodes={
                fill=white,
                text=red,
                font=\bfseries
            }
        }       
    }
}

\begin{document}

\title{\texttt{cf2vec}: Collaborative Filtering algorithm selection using graph distributed representations\thanks{This work is funded by FCT - Fundação para a Ciência e a Tecnologia - through the PhD grant SFRH/BD/117531/2016. The authors also thank the support from Brazilian funding agencies (CNPq and FAPESP), and IBM Research and Intel.}}


\titlerunning{\texttt{cf2vec}}        

\author{Tiago Cunha         \and
        Carlos Soares 		\and \\
        André C. P. L. F. de Carvalho 
}


\institute{Tiago Cunha, Carlos Soares \at
              Faculdade de Engenharia da Universidade do Porto, Portugal \\
              \email{\{tiagodscunha,csoares\}@fe.up.pt}           
           \and
           André C. P. L. F. de Carvalho \at
           Instituto de Ciências Matemáticas e de Computação - Universidade de São Paulo, Brazil \\
           \email{andre@icmc.usp.br}           
}

\date{Received: date / Accepted: date}

\maketitle

\begin{abstract}

Algorithm selection using Metalearning aims to find mappings between problem characteristics (i.e. metafeatures) with relative algorithm performance to predict the best algorithm(s) for new datasets. Therefore, it is of the utmost importance that the metafeatures used are informative. In Collaborative Filtering, recent research has created an extensive collection of such metafeatures. However, since these are created based on the practitioner's understanding of the problem, they may not capture the most relevant aspects necessary to properly characterize the problem. We propose to overcome this problem by taking advantage of Representation Learning, which is able to create an alternative problem characterizations by having the data guide the design of the representation instead of the practitioner's opinion. Our hypothesis states that such alternative representations can be used to replace standard metafeatures, hence hence leading to a more robust approach to Metalearning. We propose a novel procedure specially designed for Collaborative Filtering algorithm selection. The procedure models Collaborative Filtering as graphs and extracts distributed representations using \texttt{graph2vec}. Experimental results show that the proposed procedure creates representations that are competitive with state-of-the-art metafeatures, while requiring significantly less data and without virtually any human input. 


\keywords{Metalearning \and Deep Learning \and Representational Learning \and Collaborative Filtering \and  Recommender Systems }
\end{abstract}

\section{Introduction}

The task of recommending the best algorithms for a new given problem, also known as algorithm selection, is widely studied in Machine Learning (ML)~\citep{Brazdil2003,Prudencio2004,Smith-Miles08}. One of its most popular approaches, Metalearning (MtL), looks for a function to map metafeatures (characteristics extracted from a dataset representing the problem) to metatargets (the performance of a group of algorithms when applied to this dataset)~\citep{Brazdil2009}. This function, learned via a ML algorithm, can be used to recommend algorithms for new datasets. 

One of the main concerns in MtL is the design of metafeatures that are informative regarding algorithm performance~\citep{Vanschoren2010}. MtL has been successfully used in many ML tasks. However, since each ML task has its specificities, different sets of metafeatures may be necessary for each task. Hence, an essential part of the work of a MtL practitioner is the design of hand tailored metafeatures suitable for the task at hand. This has resulted in large collections of metafeatures for tasks like regression~\citep{Amasyali2009}, classification~\citep{Gama1995,10.1007/3-540-45357-1_26} and Collaborative Filtering~\citep{Cunha2018graphs,Cunha2018128}. 

In Collaborative Filtering (CF) algorithm selection approaches, several informative metafeatures have been proposed~\citep{Cunha2018128}, ranging from rating matrix characteristics~\citep{Adomavicius2012,Ekstrand2012,Griffith2012,Matuszyk2014,Cunha2016,Collins2018} to performance estimates on data samples~\citep{Cunha2017}. However, these metafeatures have important limitations: all are tailor made and therefore depend on he practitioner's experience and perspective on the problem. Unlike previous works, this paper investigates how useful metafeatures can be designed while minimizing human interference.

Representational Learning (RL)~\citep{Bengio2012} uses ML algorithms and domain knowledge to learn alternative and potentially richer representations for a given problem to enhance predictive performance in other ML tasks. Examples of successful applications are text classification~\citep{Bengio2012} and image recognition~\citep{He2016770}. However, to the best of our knowledge, this approach has never been used for algorithm selection tasks.

In this paper we use a RL approach to automatically design metafeatures for the problem of algorithm selection in CF. The solution proposed is inspired on distributed representations~\citep{Lecun2015}, which represent each problem entity by its underlying relationships with other lower-granularity elements. An example is \texttt{word2vec}~\citep{Mikolov2013}, which represents each word in a text using a set of neighbouring words. This paper investigates the hypothesis that there is a distributed representation technique able to create a latent representation of the CF problem, which can produce alternative metafeatures. This representation is created using \texttt{graph2vec}~\citep{Narayanan2017}, a technique inspired in \texttt{word2vec}. The proposed procedure, \texttt{cf2vec}, has 4 essential steps: 1) to convert the CF matrix into a graph, 2) to reduce the problem complexity via graph sampling, 3) to learn the distributed representations and 4) to train a metamodel with using the representation learned as metafeatures. We evaluate \texttt{cf2vec} against state-of-the-art CF metafeatures and show their performance to be comparable with state-of-the-art metafeatures, while requiring significantly less data and without virtually any human input. 

This document is organized as follows: Section~\ref{sec:rw} presents a literature review on MtL and RL for the investigated problem; Section~\ref{sec:main} introduces the proposed technique \texttt{cf2vec}; Section~\ref{sec:ex_setup} describes the experimental setup used to validate \texttt{cf2vec}, while Section~\ref{sec:results} reports the experimental analysis conducted. Finally, Section~\ref{sec:conclusions} presents the main conclusions, discusses \texttt{cf2vec}'s limitations and introduces directions for future work.

\section{Related Work}\label{sec:rw}

\subsection{Metalearning}\label{sub:sub_mtl}

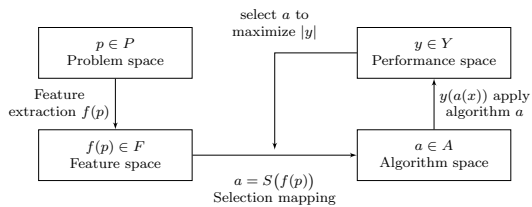
\begin{figure}[!ht]
\centering
\tikzstyle{bag} = [align=center]
 \resizebox{.6\linewidth}{!}{
  \begin{tikzpicture}[>=latex']
  \tikzset{
  block/.style= {draw, rectangle, align=center,minimum width=3cm,minimum height=1cm},
  rblock/.style={draw, rectangle, align=center,minimum width=0cm,minimum height=0cm, opacity=0},
  el/.style = {inner sep=2pt, align=left}
  }
  \node [block]  (problem) { $p \in P$ \\ Problem space};
  \node [block, below =1cm of problem] (feature) {$f(p) \in F$ \\ Feature space};        
  \node [rblock, right =1.5cm of feature] (middle) {};
  \node [block, right =1.5cm of middle] (algorithm) {$a \in A$ \\ Algorithm space};
  \node [block, above =1cm of algorithm] (performance) {$y \in Y$ \\ Performance space};
  \path[draw,->] (problem) edge node[bag] [left] {Feature \\ extraction $f(p)$} (feature) 
  (feature) edge node[bag] [below=0.25cm] {$a = S\big(f(p)\big)$ \\ Selection mapping}  (algorithm)
  (algorithm) edge node[bag] [right] {$y(a(x))$ apply \\ algorithm $a$} (performance)
  ;
  \draw[->](performance) -| node[bag] [above=0.15] {select $a$ to \\ maximize $|y|$}  (middle);
  \end{tikzpicture}
  }
\caption{Rice's Algorithm Selection conceptual framework~\citep{Smith-Miles08}).}
\label{fig:alg_sel}
\end{figure}

MtL attempts to model algorithm performance in terms of problem characteristics~\citep{Vanschoren2010}. One of its main applications is algorithm selection, first conceptualized by~\citep{DBLP:journals/ac/Rice76}.  Figure~\ref{fig:alg_sel} presents the conceptualized framework, which defines several search spaces: problem, feature, algorithm and performance, represented by $P$, $F$, $A$ and $Y$. A problem is described as: for a given instance  $ p \in P $, with features $ f(p) \in F $, find the mapping $ S(f(p)) $ into $ A $, such that the selected algorithm  $ a \in A $ maximizes $ y(a(p)) \in Y $~\citep{DBLP:journals/ac/Rice76}. Hence,  algorithm selection can be formulated as a learning task whose goal is to learn a metamodel able to recommend 
algorithms for a new task.

The first algorithm selection approaches for CF appeared recently~\citep{Adomavicius2012,Ekstrand2012,Griffith2012,Matuszyk2014}, but had low representativeness: the experimental setup and nature and diversity of metafeatures were very limited. More advanced and extensively validated CF metafeatures have been proposed:

\begin{itemize}
\item \textbf{Rating Matrix metafeatures}~\citep{Cunha2016}: these characteristics describe several rating matrix properties using a systematic metafeature generation framework~\citep{Pinto2016}. This collection of metafeatures combines sets of objects, functions and post-functions. The metafeatures proposed use three objects (rating matrix and its rows and columns), four functions (original ratings, number of ratings, mean rating value and sum of ratings) and eleven post-functions (maximum, minimum, mean, standard deviation, median, mode, entropy, gini, skewness and kurtosis). 
\item \textbf{Subsampling landmarkers}~\citep{Cunha2017}: these metafeatures are created using performance estimates on random samples extracted from the original datasets. First, random samples are extracted for each CF dataset. Next, CF algorithms are trained on these samples and their performance assessed using different evaluation measures. The outcome is a subsampling landmarker for each pair algorithm/evaluation measure. 
\item \textbf{Graph metafeatures}~\citep{Cunha2018graphs}: this approach models CF as a graph and takes advantage of the systematic metafeature extraction procedure~\citep{Pinto2016} and the hierarchical decomposition of complex data structures~\citep{Cunha:2017:MCF:3109859.3109899}. This allows to define important levels (graph, node, pairwise and subgraph) to be characterized using Graph Theory characteristics.
\item \textbf{Comprehensive metafeatures}~\citep{Cunha2018graphs}: This collection aggregates all metafeatures from all previous approaches. Correlation Feature Selection is used to obtain the most significant metafeatures. 
\end{itemize}

\subsection{Representational Learning}

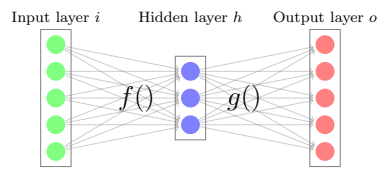
\begin{figure}[!ht]
\centering
\def\layersep{2.5cm}
\resizebox{.5\linewidth}{!}{
  \begin{tikzpicture}[shorten >=1pt,->,draw=black!50, node distance=\layersep]
      \tikzstyle{every pin edge}=[<-,shorten <=1pt]
      \tikzstyle{neuron}=[circle,fill=black!25,minimum size=10pt,inner sep=0pt]
      \tikzstyle{input neuron}=[neuron, fill=green!50];
      \tikzstyle{output neuron}=[neuron, fill=red!50];
      \tikzstyle{hidden neuron}=[neuron, fill=blue!50];
      \tikzstyle{annot} = [text width=10em, text centered]
      \tikzstyle{annot1} = [text width=10em, text centered]

      \foreach \name / \y in {1,...,5}
          \node[input neuron] (I-\name) at (0,-0.5*\y) {};

      \foreach \name / \y in {1,...,3}
          \path[yshift=-0.5cm]
              node[hidden neuron] (H-\name) at (\layersep,-0.5*\y cm) {};

      \foreach \name / \y in {1,...,5}
          \path[yshift=0cm]
              node[output neuron] (O-\name) at (2*\layersep,-0.5*\y cm) {};

      \foreach \source in {1,...,5}
          \foreach \dest in {1,...,3}
              \path[opacity=0.4] (I-\source) edge (H-\dest);

      \foreach \source in {1,...,3}
          \foreach \dest in {1,...,5}
              \path[opacity=0.4] (H-\source) edge (O-\dest);


      \node[annot,above of=H-1, node distance=1cm] (hl) {Hidden layer $h$};
      \node[annot,left of=hl] {Input layer $i$};
      \node[annot,right of=hl] {Output layer $o$};

      \node[annot1,left of=H-2, node distance=1cm] (hl) {\Large \textbf{$f()$}};
      \node[annot1,right of=H-2, node distance=1cm] (hl) {\Large \textbf{$g()$}};

       \node[draw,fit=(O-1) (O-2) (O-3) (O-4) (O-5)] {};
       \node[draw,fit=(I-1) (I-2) (I-3) (I-4) (I-5)] {};
       \node[draw,fit=(H-1) (H-2) (H-3)] {};

  \end{tikzpicture}
}
\caption{Autoencoder architecture. Autoencoders are obtained by training a neural network to reproduce the input vector in the output vector using a hidden layer with less neurons than the output layer. For such, the network learns two functions: an encoding function $f$ and a decoding function $g$. Since this hidden layer is able to preserve useful properties of the data, it can represent the input~\citep{Goodfellow-et-al-2016,Lecun2015,Schmidhuber2015}.}
\label{fig:autoencoder}
\end{figure}

Although there are alternatives, like probabilistic models and manifold learning~\citep{Bengio2011,Bengio2012}, the classical RL technique is the Autoencoder~\citep{Bourlard1988,37f2b6bee745402aa4e4d124d33be0e0}. Figure~\ref{fig:autoencoder} shows its architecture, simplified for easier readability.

Autoencoders can, theoretically, be used in any ML task, including CF. In fact, they have been used to provide recommendations~\citep{Sedhain2015,DBLP:journals/corr/StrubMG16a,Wu2016}. These works learn latent representations for each user and/or item, which are in turn used to make the recommendations. However, \texttt{cf2vec} needs a latent representation able to describe the entire dataset, like the metafeatures do. Hence, these are not useful to our purposes. 

A better alternative is the distributed representations~\citep{Lecun2015}. As the name suggests, each entity is represented by a pattern of activity distributed over many elements, and each element participates in the representation of many different entities~\citep{Rumelhart:1986:PDP:104279}. In essence, they also represent the input as a real-valued vector, but using a different network architecture. The most significant techniques for our problem are discussed next.

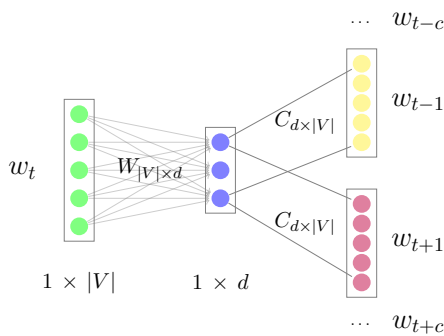
\begin{figure}[!ht]
\centering
\def\layersep{2.5cm}
\resizebox{.65\linewidth}{!}{
\begin{tikzpicture}[shorten >=1pt,->,draw=black!50, node distance=\layersep]
    \tikzstyle{every pin edge}=[<-,shorten <=1pt]
    \tikzstyle{neuron}=[circle,fill=black!25,minimum size=9pt,inner sep=0pt]
    \tikzstyle{input neuron}=[neuron, fill=green!50];
    \tikzstyle{output neuron}=[neuron, fill=red!50];
    \tikzstyle{output1 neuron}=[neuron, fill=yellow!50];
    \tikzstyle{output2 neuron}=[neuron, fill=purple!50];
    \tikzstyle{hidden neuron}=[neuron, fill=blue!50];
    \tikzstyle{annot} = [text width=10em, text centered]
	\tikzstyle{annot1} = [text width=10em, text centered]

    \foreach \name / \y in {1,...,5}
        \path[yshift=0cm]
            node[input neuron] (I-\name) at (0,-0.5*\y) {};

    \foreach \name / \y in {1,...,3}
        \path[yshift=-0.5cm]
            node[hidden neuron] (H-\name) at (\layersep,-0.5*\y cm) {};

	\foreach \name / \y in {1,...,5}
        \path[yshift=0.75cm]
            node[output1 neuron] (O-\name) at (2*\layersep,-0.35*\y cm) {};
            
    \foreach \name / \y in {6,...,10}
        \path[yshift=-0cm]
            node[output2 neuron] (O-\name) at (2*\layersep,-0.35*\y cm) {};

    \foreach \source in {1,...,5}
        \foreach \dest in {1,...,3}
            \path[opacity=0.4] (I-\source) edge (H-\dest);
            

	\path (H-1) edge[-]  (O-1);
    \path (H-3) edge[-]  (O-5);
    \path (H-1) edge[-]  (O-6);
    \path (H-3) edge[-]  (O-10);


    
    \node[annot1,below of=O-10, node distance=0.75cm] (l1) {$\dots$};
    \node[annot1,above of=O-1, node distance=0.75cm] (l2) {$\dots$};
    
    \node[annot1,left of=I-3, node distance=1cm] (hl) {\Large \textbf{$w_t$}};
    \node[annot1,right of=O-3, node distance=1cm] (hl) {\Large \textbf{$w_{t-1}$}};
    \node[annot1,right of=O-8, node distance=1cm] (hl) {\Large \textbf{$w_{t+1}$}};
    \node[annot1,right of=l1, node distance=1cm] (hl) {\Large \textbf{$w_{t+c}$}};
    \node[annot1,right of=l2, node distance=1cm] (hl) {\Large \textbf{$w_{t-c}$}};

     \node[annot1,below of=I-5, node distance=1cm] (hl) {\large \textbf{$1 \times |V|$}};
     \node[annot1,below of=H-3, node distance=1.5cm] (hl) {\large \textbf{$1 \times d $}};
     \node[annot1,left of=H-2, node distance=1.25cm] (hl) {\large \textbf{$W_{|V| \times d} $}};
     \node[annot1,left of=O-4, node distance=1cm] (hl) {\large \textbf{$C_{d \times |V|} $}};
     \node[annot1,left of=O-7, node distance=1cm] (hl) {\large \textbf{$C_{d \times |V|} $}};
     
     \node[draw,fit=(O-1) (O-2) (O-3) (O-4) (O-5)] {};
     \node[draw,fit=(O-6) (O-7) (O-8) (O-9) (O-10)] {};
     \node[draw,fit=(I-1) (I-2) (I-3) (I-4) (I-5)] {};
     \node[draw,fit=(H-1) (H-2) (H-3)] {};
     
   \end{tikzpicture}
}
\caption{Skipgram architecture used in \texttt{word2vec}~\citep{Mikolov2013}. Each target word $w_t$, represented as one-hot encoding for a vocabulary $V$, is connected to a hidden layer $h$. This hidden layer, where the distributed representations are, has a predefined size $d$. Each distributed representation is connected to the previous and next $c$ context words (i.e. $w_{t-c},w_{\dots},w_{t-1},w_{t+1},w_{\dots},w_{t+c}$). The network weights are updated until a learning stop criterion is reached.}
\label{fig:skipgram}
\end{figure}

\texttt{word2vec}~\citep{Mikolov2013} assumes that two words are similar (and have similar representations) if they have similar contexts. In this case, the context refers to a predefined amount of neighboring words. One architecture proposed to learn these representations is the skipgram, which predicts surrounding words given the current word. Figure~\ref{fig:skipgram} shows how skipgram works.

\texttt{doc2vec}~\citep{Le:2014:DRS:3044805.3045025} learns distributed representations for sequence of words with different lengths (i.e. paragraphs, documents, etc.). One of the introduced algorithms (i.e. Paragraph Vector Distributed Bag of Words (PV-DBOW)) allows a straightforward adaptation of \texttt{word2vec}'s skipgram: instead of predicting context words based on a current word, now the neural network predicts sequences of words belonging to a particular document. A variation of this technique is available in  \texttt{graph2vec}~\citep{Narayanan2017}: by considering each graph as a document, it is able to represent each graph by its underlying nodes. The process has two stages: 1) create rooted subgraphs in order to generate vocabulary and 2) train the PV-DBOW skipgram model. This technique will be discussed in detail in Section~\ref{ssub:g2v}.

\section{Distributed Representations as CF metafeatures}\label{sec:main}

This section introduces the main contribution of this work: \texttt{cf2vec}. Next, its essential steps are presented: 1) to convert the CF matrix into a graph, 2) to reduce the problem complexity via graph sampling, 3) to learn the distributed representations and 4) to train a metamodel with alternative metafeatures.

\subsection{Convert CF matrix into graph}

\begin{figure}[!ht]
    \centering
    \begin{adjustbox}{minipage=\linewidth,scale=0.85}
    \centering
    \begin{subfigure}[b]{0.45\textwidth}
    \centering
        \begin{tikzpicture}
            \matrix (first) [table0,text width=2em]
            {
            |[mylabel]| & ${\boldsymbol{i}}_{\boldsymbol{1}}$   & ${\boldsymbol{i}}_{\boldsymbol{2}}$ & ${\boldsymbol{i}}_{\boldsymbol{3}}$ \\
            $\boldsymbol{u}_{\boldsymbol{1}}$   	& ${\boldsymbol{5}}$ & ${\boldsymbol{3}}$ & ${\boldsymbol{4}}$ \\
            $\boldsymbol{u}_{\boldsymbol{2}}$    	& ${\boldsymbol{4}}$ & $\dots$ & ${\boldsymbol{2}}$  \\
            $\boldsymbol{u}_{\boldsymbol{3}}$   	& $\dots$ & ${\boldsymbol{3}}$  & ${\boldsymbol{5}}$  \\
            };
        \end{tikzpicture}
        \caption{Rating Matrix. Rows represent users $U$, while columns represent items $I$. Some cells have the rating assigned by an user to an item.}
        \label{fig:rm}
    \end{subfigure}
    ~ 
    \begin{subfigure}[b]{0.45\textwidth}
    \centering
    	\begin{tikzpicture}[thick,
                fsnode/.style={},
                ssnode/.style={},
                every fit/.style={ellipse,draw,inner sep=1pt,text width=1cm},
                ->,shorten >= 1pt,shorten <= 1pt
                ]

        \begin{scope}[start chain=going below,node distance=3mm]
        \foreach \i/\xcoord/\ycoord in {1/6/8,2/5/1,3/-4/7}
        \node[fsnode,on chain,label=left:$u_{\i}$] (f\i) {};
        \end{scope}

        \begin{scope}[xshift=3cm,start chain=going below,node distance=3mm]
        \foreach \i/\xcoord/\ycoord in {1/0/3,2/1/4,3/-2/1}
        \node[ssnode,on chain,label=right:$i_{\i}$] (s\i) {};
        \end{scope}

        \node [myblue,fit=(f1) (f3),label=above:$U$, line width=0.5mm] {};
        \node [mygreen,fit=(s1) (s3),label=above:$I$, line width=0.5mm] {};

        \draw (f1) edge[bend left=45] node[label={[xshift=0cm,yshift=-0.05cm]5}]{} (s1) ;
        \draw (f1) edge[bend left=30] node[label={[xshift=-0.1cm,yshift=-0.05cm]3}]{} (s2);
        \draw (f1) edge[bend left=15] node[label={[xshift=-0.1cm,yshift=-0.5cm]4}]{}  (s3);
        \draw (f2) edge[bend left=15] node[label={[xshift=0.02cm,yshift=-0.45cm]4}]{} (s1);
        \draw (f2) edge[bend right=15] node[label={[xshift=0.04cm,yshift=-0.2cm]2}]{} (s3);
        \draw (f3) edge[bend right=30] node[label={[xshift=-0.05cm,yshift=-0.5cm]3}]{} (s2);
        \draw (f3) edge[bend right=45] node[label={[xshift=0.03cm,yshift=-0.6cm]5}]{} (s3);
        \end{tikzpicture}
        \caption{Bipartite Graph. The graph has two node subsets, representing users $U$ and items $I$. Ratings are  weighted edges between nodes of both subsets.}
        \label{fig:gr}
    \end{subfigure}
    \end{adjustbox}
    \caption{Toy example for two different CF representations.}\label{fig:toy_example}
\end{figure}
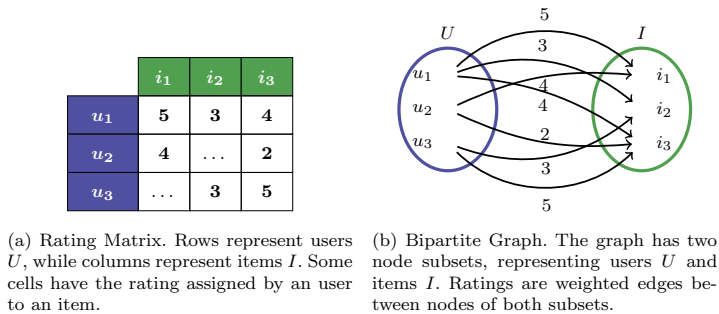

CF is usually described by a rating matrix $R ^{|U| \times |I|}$, representing a set of users $U$ and items $I$. Each element of this matrix is the feedback provided by each user for each item. Figure~\ref{fig:rm} shows a toy example of a rating matrix.

To use \texttt{graph2vec}, the input elements must be graphs. Since~\citep{Cunha2018graphs} have shown that a CF rating matrix can be seen as an adjacency matrix, then the problem can be stated as: consider a bipartite graph $G$, whose nodes $U$ and $I$ represent users and items, respectively. The edges $E$ connects elements of the two groups and represent the feedback provided by users to items. The edges can be weighted in order to represent preference values (ratings). Figure~\ref{fig:gr} shows the conversion of the toy example from Figure~\ref{fig:rm}. 

\subsection{Sampling graphs}

An important part of metafeature design is the effort required~\citep{Vanschoren2010}: if the task is slower than training and evaluating all algorithms on the new problem, then it is useless. Considering how CF graphs can reach quite large sizes, this is a pressing issue and it motivates our necessity in reducing the problem dimensionality. Since one is not interested in the actual time required, but rather on reducing the amount of data to be processed in order to reduce the time needed, the focus lies on investigating which is the minimum  amount of data which allows to maintain a high predictive performance. 

Thus, an intermediate (but not mandatory) step is added: graph sampling. In order to find a distributed representation as closely related as possible to the entire graph, a sampling technique able to preserve the graph structural properties must be chosen. According to~\citep{Leskovec2006}, a good choice is random walk. It performs multiple explorations of graph paths until $\theta$ nodes are reached and uses all of them to obtain the respective subgraph. 

\subsection{Learn distributed representation}

Taking advantage of \texttt{graph2vec}'s agnostic nature, one argues that the problem can be defined as follows: given a set of CF graphs $G = \{g_1, g_2, ...\}$ and a positive integer $\sigma$ (i.e., distributed representation size), one aims to learn a $\sigma$-dimensional distributed representation for every graph. Hence, this process creates a matrix of distributed representations $E^{|G| \times \sigma}$, which can be regarded as a metafeature representations for all considered graphs. Two steps are required: 1) to extract of rooted subgraphs and 2) to learn matrix $E$.

\subsubsection{Extract rooted subgraphs}

A rooted subgraph $sg_{n}^{\delta}$ is composed by the set of nodes (and corresponding edges) around node $n \in g_i$ that are reachable in $\delta$ hops. Learning the distributed representation requires the extraction of rooted subgraphs for all nodes. Thus, the process must be applied to $N$ nodes, in which $N = |U| + |I|$. 

Rooted subgraphs in \texttt{graph2vec} are generated using the Weisfeiler-Lehman relabeling procedure~\citep{Shervashidze:2011:WGK:1953048.2078187}. Beyond being able to inspect neighboring nodes, it is also able to incorporate information about the neighbors in a single node's name. As a result, it creates a rich textual description for every graph. To do so, it iteratively traverses each original node and using all neighbors as the current node label. Next, it replaces the original node labels by new compressed names, which represent a neighborhood structure. The process repeats until $d$ hops are reached. Every rooted subgraph can be represented by a numeric vector with the frequency each node (original or compressed) appears in the representation, similar to one-hot encoding. 

\subsubsection{Learn matrix $E$}\label{ssub:g2v}

Considering how now there is a graph vocabulary, then the skipgram model can be used straightforwardly. As it can be seen in Figure~\ref{fig:skipgram_g2v}, each graph $g_i$ is represented by its identifier and connected to $\delta$ context rooted subgraphs $sg$. Training such a neural network allows to learn similar distributed representations for graphs with similar rooted subgraphs. The authors believe that this relationship also relates with algorithm performance, similarly to what happens in other metafeatures. Hence, it is suitable for algorithm selection.

\begin{figure}[!ht]
\centering
\def\layersep{2.5cm}
\resizebox{.55\linewidth}{!}{
\begin{tikzpicture}[shorten >=1pt,->,draw=black!50, node distance=\layersep]
    \tikzstyle{every pin edge}=[<-,shorten <=1pt]
    \tikzstyle{neuron}=[circle,fill=black!25,minimum size=9pt,inner sep=0pt]
    \tikzstyle{input neuron}=[neuron, fill=green!50];
    \tikzstyle{output neuron}=[neuron, fill=red!50];
    \tikzstyle{output1 neuron}=[neuron, fill=yellow!50];
    \tikzstyle{output2 neuron}=[neuron, fill=purple!50];
    \tikzstyle{hidden neuron}=[neuron, fill=blue!50];
    \tikzstyle{annot} = [text width=10em, text centered]
	\tikzstyle{annot1} = [text width=10em, text centered]

    \foreach \name / \y in {1,...,1}
        \path[yshift=0cm]
            node[input neuron] (I-\name) at (0,-1.5*\y) {};

    \foreach \name / \y in {1,...,3}
        \path[yshift=-0.5cm]
            node[hidden neuron] (H-\name) at (\layersep,-0.5*\y cm) {};

	\foreach \name / \y in {1,...,5}
        \path[yshift=0.75cm]
            node[output1 neuron] (O-\name) at (2*\layersep,-0.35*\y cm) {};
            
    \foreach \name / \y in {6,...,10}
        \path[yshift=-0cm]
            node[output2 neuron] (O-\name) at (2*\layersep,-0.35*\y cm) {};

            

	\path (I-1) edge[-]  (H-1);
    \path (I-1) edge[-]  (H-3);
	\path (H-1) edge[-]  (O-1);
    \path (H-3) edge[-]  (O-5);
    \path (H-1) edge[-]  (O-6);
    \path (H-3) edge[-]  (O-10);


    
    \node[annot1,below of=O-4, node distance=.9cm] (l1) {$\dots$};
    
    \node[annot1,left of=I-3, node distance=1cm] (hl) {\Large \textbf{$g_i$}};
    \node[annot1,right of=O-3, node distance=1cm] (hl) {\Large \textbf{$sg_{1}$}};
    \node[annot1,right of=O-8, node distance=1cm] (hl) {\Large \textbf{$sg_{\delta}$}};
    \node[annot1,right of=l1, node distance=1cm] (hl) {\Large \textbf{$\dots$}};

    \node[annot1,left of=H-2, node distance=1cm] (hl) {\large \textbf{$E^{|G| \times \sigma}$}};
      \node[annot1,left of=O-4, node distance=1cm] (hl) {\large \textbf{$C_{\sigma \times N} $}};
      \node[annot1,left of=O-7, node distance=1cm] (hl) {\large \textbf{$C_{\sigma \times N} $}};
     
     \node[draw,fit=(O-1) (O-2) (O-3) (O-4) (O-5)] {};
     \node[draw,fit=(O-6) (O-7) (O-8) (O-9) (O-10)] {};
     \node[draw,fit=(I-1)] {};
     \node[draw,fit=(H-1) (H-2) (H-3)] {};
     
   \end{tikzpicture}
}
\caption{Skipgram architecture used in \texttt{graph2vec}~\citep{Narayanan2017}. }
\label{fig:skipgram_g2v}
\end{figure}

In order to learn the weights, then one must train the network. The learning process, based on Stochastic Gradient Descent, iteratively performs these steps until conversion is achieved: 1) feedforward weights from input to the output layer, 2) application of a softmax classifier to compare the output layer's weights with the subgraph representations and 3) backpropagation of the errors through the network. Doing so, it learns matrices $E$ and $C$, which represent the distribuetd representations and context matrices, respectively. Notice the skipgram is trained using Negative Sampling, which does not use all subgraphs belong to a graph. Instead, it takes advantage of few random subgraphs that do not belong to the graph. This way, training is more efficient.

\subsection{Learn metamodel}

Notice that matrix $E$ can be easily used as metafeatures. Thus, every problem $p_i$ is described by independent variables (the $i$-th row of matrix $E$) and the dependent variables (the respective ranking of algorithms). Obtaining these pairs for all $g_i$, allows to create a metadatabase like the one in Figure~\ref{fig:mtl_process}. 

\begin{figure}[!ht]
\centering
\begin{adjustbox}{minipage=\linewidth,scale=0.85}
\centering
\begin{tikzpicture}
\matrix (first) [table,text width=3.5em]
{
$\boldsymbol{P}$ & ${\boldsymbol{f}}_{\boldsymbol{1}}\boldsymbol{()}$  & $\dots$ & ${\boldsymbol{f}}_{\boldsymbol{|F|}}\boldsymbol{()}$ \\
$\boldsymbol{p}_{\boldsymbol{1}}$   	& ${\boldsymbol{\omega}}_{\boldsymbol{1}}$ & $\dots$ & ${\boldsymbol{\omega}}_{\boldsymbol{|F|}}$ \\
$\vdots$   	& $\vdots$ & $\ddots$ & $\vdots$ \\
$\boldsymbol{p}_{\boldsymbol{|P|}}$   	& $\dots$ & $\dots$ & $\dots$ \\
};

\matrix (second) [table2,text width=2.5em, right=0.5cm of first]
{
${\boldsymbol{a}}_{\boldsymbol{1}}$ & $\dots$ & ${\boldsymbol{a}}_{\boldsymbol{|A|}}$\\
$\boldsymbol{\pi}_{\boldsymbol{1}}$ & $\dots$ & $\boldsymbol{\pi}_{\boldsymbol{|A|}}$\\
$\vdots$ & $\ddots$ & $\vdots$ \\
$\dots$ & $\dots$ & $\dots$ \\
};

\matrix (third) [table3,text width=3.5em, below=0.25cm of first]
{
$\boldsymbol{p}_{\boldsymbol{\alpha}}$  & ${\boldsymbol{\hat{\omega}}}_{\boldsymbol{1}}$ & $\dots$ & ${\boldsymbol{\hat{\omega}}}_{\boldsymbol{|F|}}$\\
};

\matrix (fourth) [table4,text width=2.5em, below=0.25cm of second]
{
$\hat{\boldsymbol{\pi}}_{\boldsymbol{1}}$ & $\boldsymbol{\dots}$ & $\hat{\boldsymbol{\pi}}_{{\boldsymbol{|A|}}}$\\
};

\end{tikzpicture}
\end{adjustbox}
\caption{Metadatabase. Organized into training and prediction data (top, bottom) and independent and dependent variables (left, right).}
\label{fig:mtl_process}
\end{figure}

Formally, the submission of all problems $p_i$ (i.e. $g_i$) to \texttt{cf2vec} produces the metafeatures $\omega = f(p_i)$. To create the dependent variables, each problem $p_i$ is associated with the respective ranking of algorithms $\pi$, based on the performance values for a specific evaluation measure $y_k \in Y$. This ranking considers a static ordering of the algorithms $a_j$ (using for instance an alphabetical order) and is composed by a permutation of values $\{1,...,|A|\}$. These values indicate, for each position $l$, the respective ranking. A learning algorithm is then used to induce a metamodel. In order to make predictions, the metamodel can be applied to metafeatures $\hat{\omega} = {f}(p_{\alpha})$ extracted from a new problem $p_{\alpha}$ to predict its best ranking of algorithms $\hat{\pi}$. 

Considering how the problem is modelled, the ideal solution is Label Ranking (LR)~\citep{Hullermeier2008,Vembu2010}. Thus, the algorithm selection problem for CF using LR is: for every dataset $p \in P$, with features $f(p) \in F$ associated with the respective rankings $\pi_p$, find the selection mapping $g(f(p))$ into the permutation space $\Omega$, such that the selected ranking of algorithms  $ \pi_p $ maximizes the performance mapping $ y(\pi_p) \in Y $.

\section{Experimental setup}\label{sec:ex_setup}

Any MtL-based problem for algorithm selection has two well defined levels: the baselevel (a conventional ML task applying ML algorithms to problem-related datasets) and the metalevel (apply ML algorithms to metadatasets). In this work, the base level is a CF task and the metalevel is the application of ML algorithms for Label Ranking. From this point onward baselearners and metalearners are the algorithms used in the baselevel and in the metalevel, respectively. Next, the experimental setup used in this work is presented. 

\subsection{Collaborative Filtering}

In the baselevel, CF baselearners are applied to CF datasets and evaluated using CF assessment measures. It uses 38 datasets, described in Table~\ref{tab:cf_data}, alongside a summary of their statistics, namely the number of users, items and ratings. Due to space restrictions, the datasets are identified by acronyms: Amazon (AMZ), Bookcrossing (BC), Flixter (FL), Jester (JT), MovieLens (ML), MovieTweetings (MT), TripAdvisor (TA), Yahoo! (YH) and Yelp (YE).

\begin{table}[!ht]	
\centering
\scriptsize
\caption{Summary dataset description.} \label{tab:cf_data}
\begin{tabular}{l|c|c|c|l}
\hline
	Dataset & \#users & \#items & \#ratings & Reference\\ \hline
	AMZ-apps  & 132391 & 24366 & 264233 & \multirow{22}{*}{\citep{McAuley2013}}\\ \cline{1-4}
	AMZ-automotive  & 85142 & 73135 & 138039 \\ \cline{1-4}
	AMZ-baby  & 53188 & 23092 & 91468 \\ \cline{1-4}
	AMZ-beauty  & 121027 & 76253 & 202719 \\ \cline{1-4}
	AMZ-cd  & 157862 & 151198 & 371275 \\ \cline{1-4}
	AMZ-clothes  & 311726 & 267503 & 574029 \\ \cline{1-4}
	AMZ-food  & 76844 & 51139 & 130235 \\ \cline{1-4}
	AMZ-games  & 82676 & 24600 & 133726 \\ \cline{1-4}
	AMZ-garden  & 71480 & 34004 & 99111 \\ \cline{1-4}
	AMZ-health  & 185112 & 84108 & 298802 \\ \cline{1-4}
	AMZ-home  & 251162 & 123878 & 425764 \\ \cline{1-4}
	AMZ-instruments  & 33922 & 22964 & 50394 \\ \cline{1-4}
	AMZ-kindle  & 137107 & 131122 & 308158 \\ \cline{1-4}
	AMZ-movies  & 7278 & 1847 & 11215 \\ \cline{1-4}
    AMZ-music  & 47824 & 47313 & 83863 \\ \cline{1-4}
	AMZ-office  & 90932 & 39229 & 124095 \\ \cline{1-4}
	AMZ-pet Supplies  & 74099 & 33852 & 123236 \\ \cline{1-4}
	AMZ-phones  & 226105 & 91289 & 345285 \\ \cline{1-4}
	AMZ-sports  & 199052 & 127620 & 326941 \\ \cline{1-4}
	AMZ-tools  & 121248 & 73742 & 192015 \\ \cline{1-4}
	AMZ-toys  & 134291 & 94594 & 225670 \\ \cline{1-5}
    AMZ-video  & 42692 & 8882 & 58437 \\ \cline{1-4}
	BC  & 7780 & 29533 & 39944 & \multirow{1}{*}{\citep{Ziegler2005}}\\ \cline{1-5}
	FL  & 14761 & 22040 & 812930 & \multirow{1}{*}{\citep{Zafarani+Liu:2009}}\\ \cline{1-5}
	JT1  & 2498 & 100 & 181560 & \multirow{3}{*}{\citep{Goldberg2001}}\\ \cline{1-4}
	JT2  & 2350 & 100 & 169783 \\ \cline{1-4}
	JT3  & 2493 & 96 & 61770 \\ \cline{1-5}
	ML100k  & 94 & 1202 & 9759 & \multirow{5}{*}{\citep{GroupLens2016} }\\ \cline{1-4}
	ML10m  & 6987 & 9814 & 1017159 \\ \cline{1-4}
	ML1m  & 604 & 3421 & 106926 \\ \cline{1-4}
	ML20m  & 13849 & 16680 & 2036552 \\ \cline{1-4}
	ML-latest  & 22906 & 17133 & 2111176 \\ \cline{1-5}
	MT-latest  & 3702 & 7358 & 39097 & \multirow{2}{*}{\citep{Dooms13crowdrec}}\\ \cline{1-4}
	MT-RS14  & 2491 & 4754 & 20913 \\ \cline{1-5}
	TA  & 77851 & 10590 & 151030 & \multirow{1}{*}{\citep{Wang2011} }\\ \cline{1-5}
	YH-movies  & 764 & 4078 & 22135 & \multirow{2}{*}{\citep{Yahoo} }\\ \cline{1-4}
	YH-music  & 613 & 4620 & 30852 \\ \cline{1-5}
	YE  & 55233 & 46045 & 211627 & \multirow{1}{*}{\citep{Yelp2016}}\\ \cline{1-5}
\end{tabular}
\end{table}

The experiments were carried out with MyMediaLite~\citep{Gantner2011}. Two CF tasks were addressed: Rating Prediction and Item Recommendation. While the first aims to predict the rating an user would assign to a new instance, the second aims to recommend a ranked list of items. Since the tasks are different, so are the baselearners and evaluation measures required. 

The following CF baselearners were used for Rating Prediction: Matrix Factorization (MF), Biased MF (BMF)~\citep{Salakhutdinov2008}, Latent Feature Log Linear Model (LFLLM)~\citep{Menon2010}, SVD++~\citep{Koren2008}, 3 versions of Sigmoid Asymmetric Factor Model (SIAFM, SUAFM and SCAFM)~\citep{Paterek2007}, User Item Baseline (UIB)~\citep{Koren2010} and Global Average (GA). For Item Recommendation, the baselearners chosen were BPRMF~\citep{Rendle2009}, Weighted BPRMF (WBPRMF)~\citep{Rendle2009}, Soft Margin Ranking MF (SMRMF)~\citep{Weimer2008}, WRMF~\citep{Hu2008a} and Most Popular (MP). The baselearners were selected based on the fact that all are Matrix Factorization algorithms, well known for their predictive power and computational efficiency.

In the Item Recommendation experiments the baselearners were evaluated using NDCG and AUC, while for Rating Prediction, NMAE and RMSE were used. All experiments were performed using 10-fold cross-validation. To prevent bias in favour of any baselearner, the hyperparameters were not tuned.

\subsection{Metalearning with Label Ranking}

The metalevel use metalearners to map metafeatures to metatargets. This work investigates two types of metafeatures:
\begin{itemize}
\item Comprehensive metafeatures~\citep{Cunha2018graphs}: chosen to represent standard MtL approaches, since they achieve the best performance and represent the most diverse set of problem characteristics. 
\item \texttt{cf2vec} metafeatures: distributed representations learned from the proposed procedure. One important issue to address is the hyperparameter optimization since depending on their settings, different representations are produced. This work pays special attention to $\delta$ and $\sigma$, since they were shown to be the most important in~\citep{Mikolov2013}. However, all hyperparameters are tuned using grid search~\citep{BergstraJAMESBERGSTRA2012}.
\end{itemize}

The multicriteria metatargets procedure used in the latest related work approach for CF algorithm selection~\citep{Cunha2018graphs} is replicated here. The authors introduce a novel way to model the metatargets, which is able to create a single ranking of algorithms by considering more than one evaluation measure. This decision has been made since it allows to create fairer rankings and to reduce the amount of algorithm selection problems investigated.

This work uses only one metalearner, since one aims to simplify the presentation of results. To that end, KNN~\citep{Soares2015} was chosen due to its superior predictive performance in CF algorithm selection~\citep{Cunha2018graphs}. The experiments use as baseline the Average Rankings algorithm~\citep{Brazdil2000}. Metamodels are evaluated using Kendall's Tau and leave one out cross-validation and tuned using grid search~\citep{BergstraJAMESBERGSTRA2012}.

\section{Results and Discussion}\label{sec:results}

Here, a set of Research Questions (RQs) are posed to empirically compare the merits of \texttt{cf2vec}'s distributed representations against CF metafeatures. 

\begin{example1}
Which is the best $\theta$ setting in \texttt{cf2vec}?
\end{example1}

\begin{figure}[!ht]
  \centering
    \includegraphics[width=.8\linewidth]{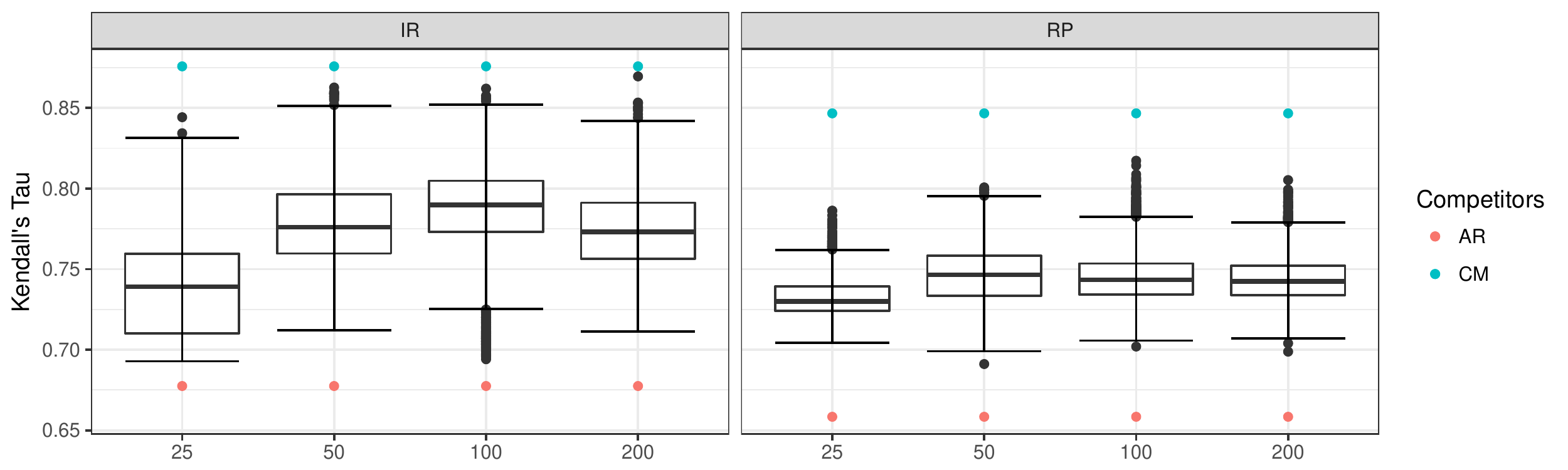}
      \caption{Kendall's tau in terms of $\theta$ (amount of nodes sampled per graph).}
      \label{fig:meta_sampling}
\end{figure}

This analysis investigates the effect of $\theta$ (amount of nodes sampled per graph) on Kendall's tau performance, which measures how similar are the true and predicted rankings of CF algorithms averaged by all datasets considered. Figure~\ref{fig:meta_sampling} shows the distribution of Kendall's tau scores for all \texttt{cf2vec} metamodels, with $\theta \in \{25,50,100,200\}$. The results also show the performance obtained with Comprehensive Metafeatures (CM) and Average Rankings (AR). 

According to these results:
\begin{itemize}
\item \texttt{cf2vec} creates informative representations: this is supported by the fact that all their performances are better than the baseline AR. 
\item \texttt{cf2vec} is never better than CM: although the performance results come very close to CM's, this threshold is never beaten. 
\item The best settings is $\theta = 100$: although the performances are quite similar, it is visible that the best and median performances increase until this threshold, but suffer a small decrease afterwards.
\end{itemize}

\begin{example1}
How \texttt{cf2vec} performance is affected by \texttt{graph2vec}'s hyperparameters?
\end{example1}

\begin{figure}[!ht]
  \centering
    \includegraphics[width=.8\linewidth]{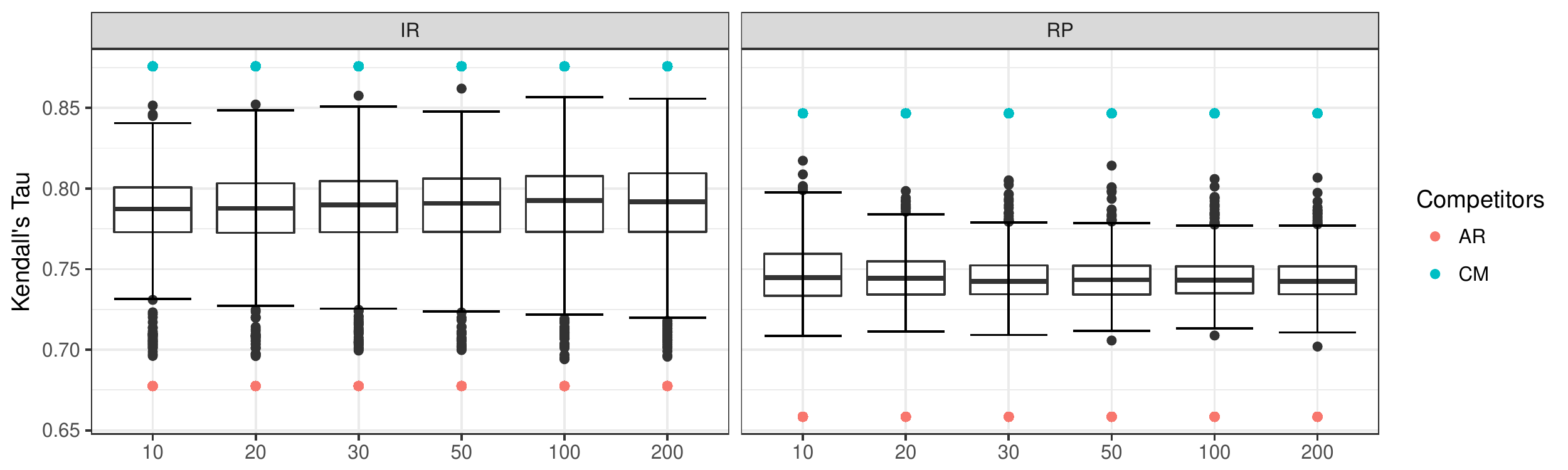}
      \caption{Kendall's tau in terms of $\sigma$ (distributed representation size).}
      \label{fig:meta_dims}
\end{figure}

\begin{figure}[!ht]
  \centering
    \includegraphics[width=.8\linewidth]{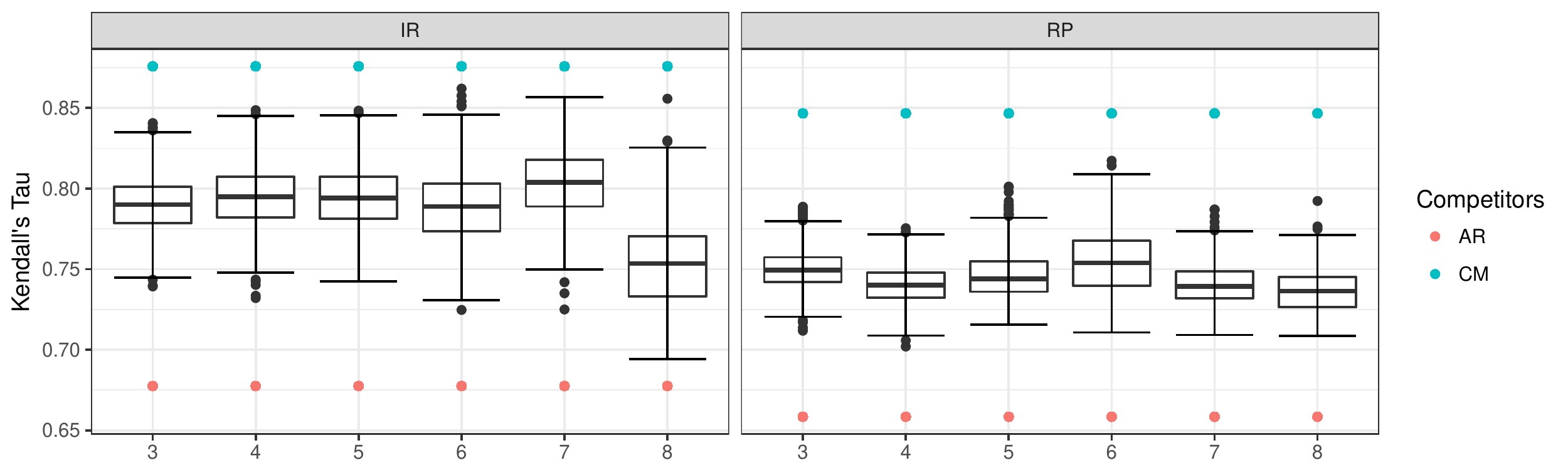}
      \caption{Kendall's tau in terms of $\delta$ (amount of context subgraphs).}
      \label{fig:meta_wl}
\end{figure}

This analysis focusses on two hyperparameters: $\sigma$ and $\delta$, the representation size and the amount of context subgraphs, respectively. All other hyperparameters are disregarded since no obvious patterns emerged. Figures~\ref{fig:meta_dims} and~\ref{fig:meta_wl} present Kendall's tau performance for all \texttt{cf2vec} metamodels built with $\theta = 100$, since this proved to be the best setting. 

According to these results:
\begin{itemize}
\item The performances for $\sigma$ are stable: although the best and worst performances slightly fluctuate, the median values remain the same. These are surprising results, which may be explained by limited grid search settings. However, considering the reduced amount of meta-examples and the curse of dimensionality, it would be difficult to improve these results.
\item Hyperparameter $\delta$ has a significant impact on the predictive performance: both metatargets increase their performance until $\delta = 6$. Soon after, their performances decreases. However, lower amounts of context subgraphs lead to better performance (look how $\delta \in \{3,4,5\}$ perform better than $\delta = 8$). 
\end{itemize}

\begin{example1}
How does the performance of the best \texttt{cf2vec} metamodel compare with the best metamodel induced with Comprehensive Metafeatures?
\end{example1}

\begin{figure}[!ht]
  \centering
    \includegraphics[width=.9\linewidth]{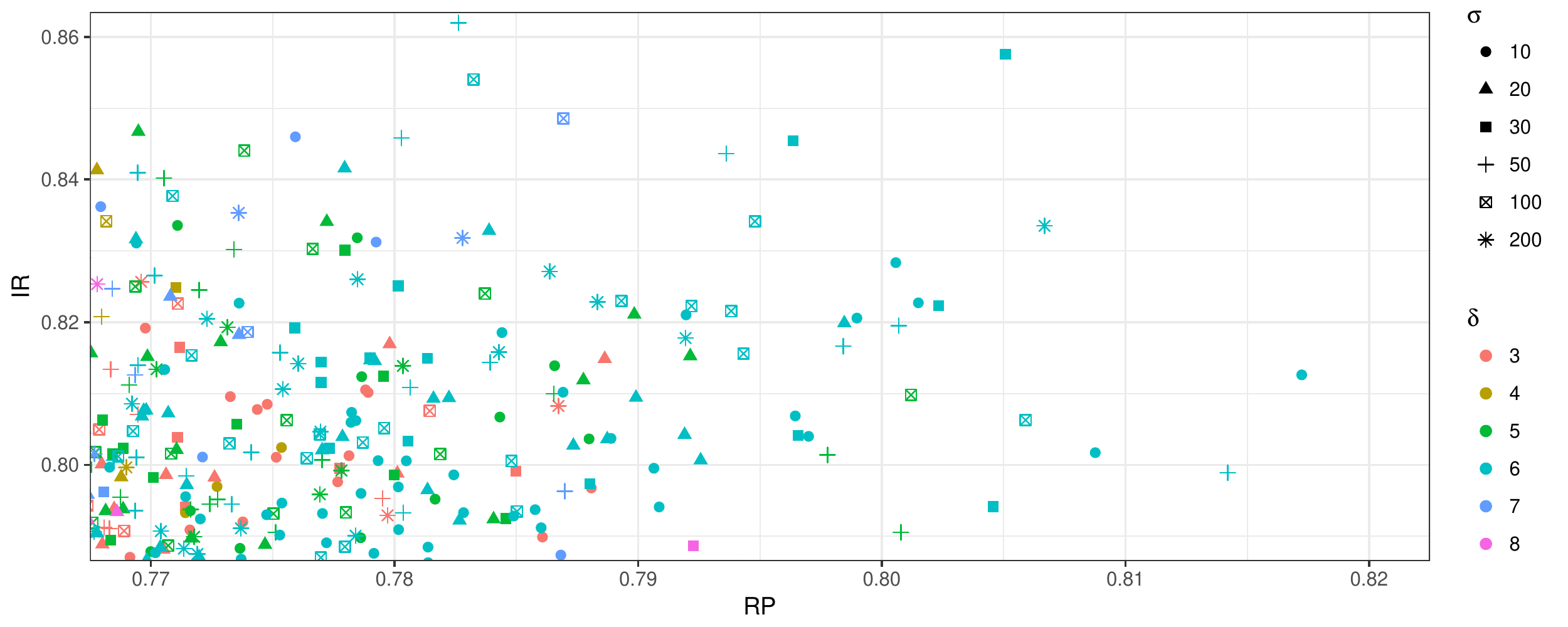}
      \caption{Performance scatterplot in both CF problems.}
      \label{fig:meta_all}
\end{figure}

To select the best \texttt{cf2vec} hyperparameter settings, the best performance on both CF problems must be found. To illustrate how the performance is distributed, Figure~\ref{fig:meta_all} presents the best Kendall's tau performances for both problems. The metamodels are identified by their $\sigma$ and $\delta$ hyperparameters. 

The results show that metamodels with $\delta=6$ occupy the vast majority of performances that simultaneously maximize the performance on both tasks. Among these, the best hyperparameter settings correspond to the performance point placed at $(0.805, 0.858)$, in which $\sigma = 30$. The metamodel trained with this hyperparameter settings is henceforth used as \texttt{cf2vec}'s representative.

\begin{figure}[!ht]
  \centering
    \includegraphics[width=.75\linewidth,trim={2.5cm 2.5cm 1cm 2cm},clip]{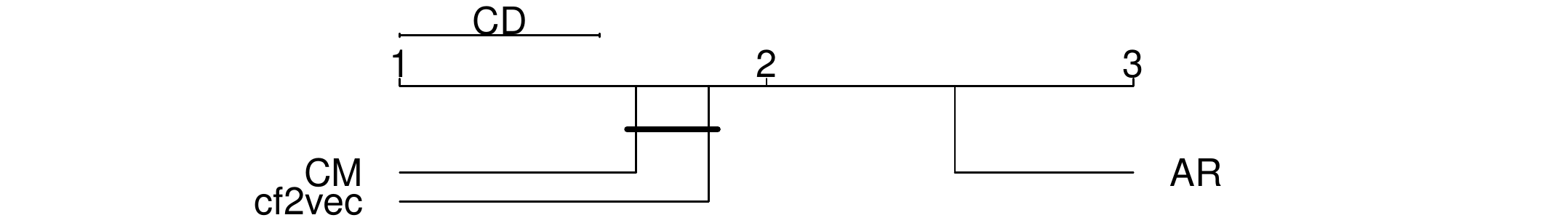}
      \caption{Critical difference diagram for the best hyperparameter settings.}
      \label{fig:cd}
\end{figure}

To understand how \texttt{cf2vec} competes with other strategies, a statistically significance test was used: Critical Difference (CD) diagrams~\citep{Demsar2006}. Each strategy is represented by its best metamodel's performances for all datasets. These are used here to rank several metalearners and to assess whether the differences are statistically significant. The CD interval created - which is calculated with a Friedman's test - connects all metalearners for which there is no statistically significant difference. Figure~\ref{fig:cd} shows these results.

The results show that there is no statistically significant difference between CM and \texttt{cf2vec} and that both are better than the baseline. Thus, the proposed approach is not only suitable to the task at hand, but it is also as good as the best CF metafeatures. 

\begin{example1}
What is the impact on the baselevel performance achieved by \texttt{cf2vec}?
\end{example1}

\begin{figure}[!ht]
  \centering
    \includegraphics[width=.75\linewidth]{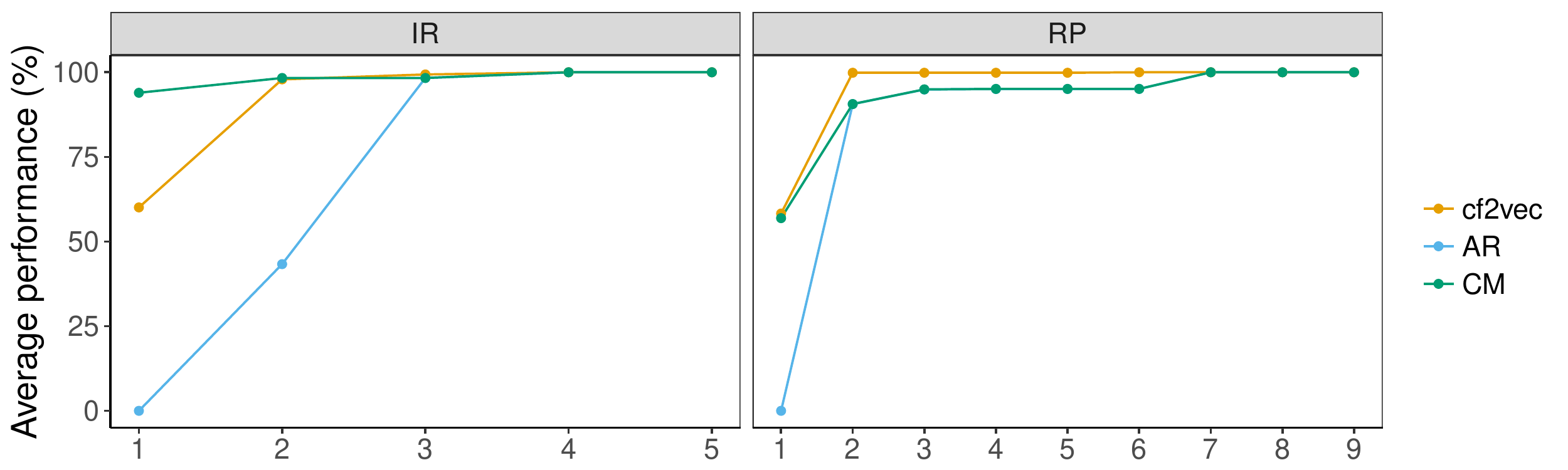}
      \caption{Impact on the baselevel performance.}
      \label{fig:base}
\end{figure}

Since differences in baselevel performances in a ranking can be quite costly, it is essential to assess each metalearner by the baselevel predictive performance of its predicted rankings. To do so, each threshold $t$ in the predicted ranking of algorithms is replaced by the respective baselearner's performance. This performance vector is then normalized and averaged by all datasets. Figure~\ref{fig:base} shows these results for both CF problems. The amount of thresholds $t$ is different because each problem has a different amount of algorithms. The results are presented in percentage in order to facilitate interpretation.

The results show that while CM is better for $t=1$ in Item Recommendation, for all remaining thresholds, \texttt{cf2vec} and CM have the same performance. However, in Rating Prediction, \texttt{cf2vec} is better than CM for $t \in \{2,3,4,5,6\}$ and equal for $t=1$. These results show that \texttt{cf2vec} obtains a comparable (and even higher) baselevel performance.

\begin{example1}
Is there a clear relation between metafeatures (either CM or \texttt{cf2vec} representations) and CF algorithm performance?
\end{example1}

Considering the similarity in predictive performance for both types of metafeatures in all previous analysis, it is important to understand whether there are clear relationships between them and the metatargets. These relations can potentialy explain the results obtained.

\begin{figure}[!ht]
  \centering
    \includegraphics[width=1\linewidth]{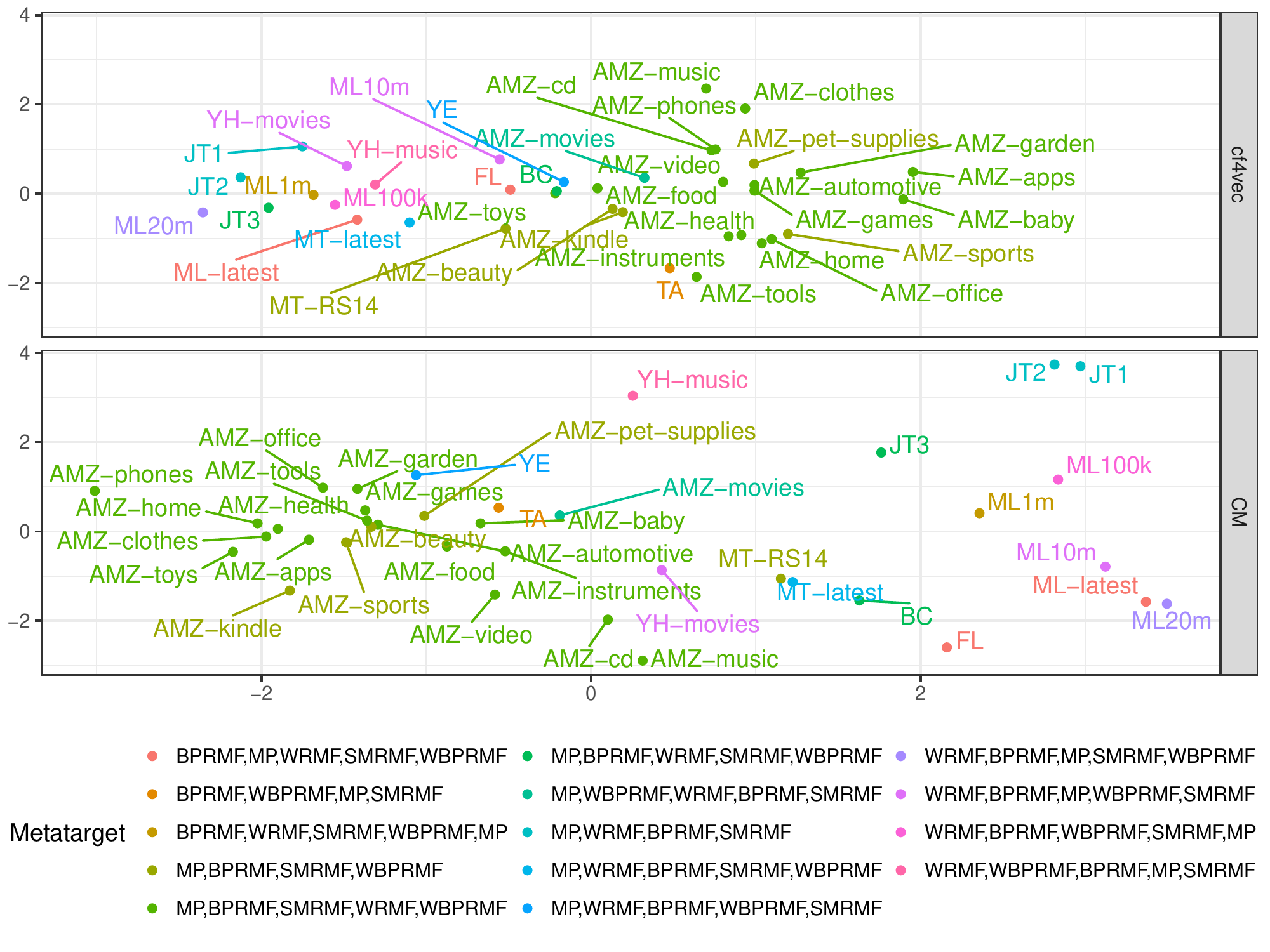}
      \caption{PCA visualization for Item Recommendation problem.}
      \label{fig:tsne_ir}
\end{figure}

\begin{figure}[!ht]
  \centering
    \includegraphics[width=1\linewidth]{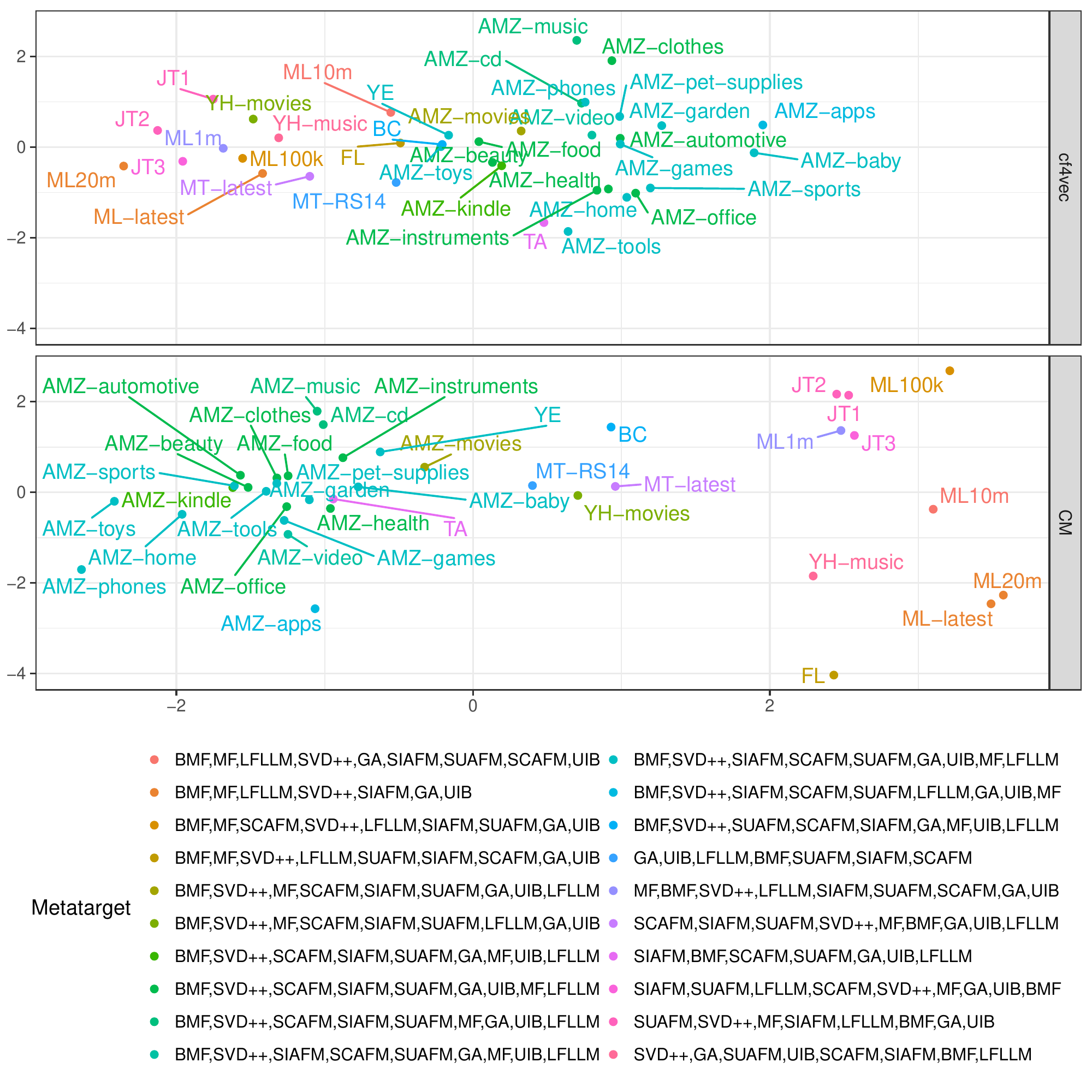}
      \caption{PCA visualization for Rating Prediction problem.}
      \label{fig:tsne_rp}
\end{figure}

The literature in distributed representations often refers to t-sne~\citep{VanDerMaaten2008} to explore high-fimensional datasets. However, two limitations have shown this technique to not be ideal in our setup: 1) the procedure is stochastic, hence the representations are not static and 2) the process, being specially designed for a large amount of data points, is not ideal to our problem, where only 38 data points exist. Hence, PCA was used to visualize the high-dimensional metafeatures in a two dimensional map. To enrich the results, the ranking of baselearners for each dataset is shown using a colour gradient which is assigned based on metatarget similarity. This highlights clear patterns between metafeatures and metatargets: if similar (or the same) metatargets are assigned to two similar datasets (placed near one another) there is a clear pattern. Figures~\ref{fig:tsne_ir} and~\ref{fig:tsne_rp} illustrate the results.

The results show that both metafeatures work well in two cases:
\begin{itemize}
\item Same domain and similar metatargets: most datasets from the same domain have clearly visible patterns in the mappings between metafeatures and metatargets. This occurs for the AMZ and JT domains.
\item Different domains but similar metatargets: some datasets from different domains, and sharing similar metatargets, are close to each other. This happens for the BC and FL datasets in Item Recommendation and for the YE and FL datasets in Rating Prediction.
\end{itemize}

The previous observations refer to the easily predictable meta-instances. The fact that both types of metafeatures are able to properly map the instances together is a good reason to explain why they perform well for the majority of datasets. However, some problems were found:

\begin{itemize}
\item Anomalies: some points are close to others without any apparent reason. This occurs in the TA dataset for both CF problems and the YE dataset in the Item recommendation problem. This may have occurred because the current metafeatures are not good enough to characterize these datasets.
\item Same domain but different metatargets: some datasets from the same domain appear close. However, their rankings are significantly different. This occurs for the ML, YH and MT datasets. A possible reason is difficulty of the metamodel to correctly predict the rankings of algorithms. This difficulty can be potentially be reduced by tuning the metalearner hyperparameters and by chosing metalearners with different bias.
\end{itemize}

The low occurrence of these problems can explain the high predictive performance obtained. However, their occurrence points out the need for further studies. One important issue lies in the need to find more and more diverse datasets and baselearners to complement such observations. 

Finally, although \texttt{cf2vec} shows interesting and useful patterns, CM seems to be generally better at mapping the difficult problems. This may be the missing indicator which justify the differences in predictive performance. Therefore, although the hypothesis of using distributed representations as metafeatures alternatives is validated, it is clear that other techniques may be needed to surpass the performance of several hand designed metafeatures specially designed for this empirical setup. Nevertheless, the research direction presented shows promising new algorithm selection solutions in CF and other ML tasks.

\section{Conclusions}\label{sec:conclusions}

This paper introduced a novel technique for CF metafeature design: \texttt{cf2vec}. \texttt{cf2vec} adapts a known distributed representation technique \texttt{graph2vec} to the context of CF algorithm selection. To do so, the procedure converts CF datasets into graphs, reduces the problem complexity via graph sampling, learns the distributed representation and uses them as alternative metafeatures. Experiments carried out show that \texttt{cf2vec} is competitive with the state-of-the-art collection of CF metafeatures, with no statistically significant differences. The main advantages of \texttt{cf2vec} are: the metafeatures are automatically generated without virtually any human intervention, the process can be tuned to adjust the metafeatures to the experimental setup and the procedure reaches the same performance as the state-of-the-art, but requiring a smaller amount of data. However, \texttt{cf2vec} also has limitations. These are discussed next, along with suggestions to deal with them:

\begin{itemize}
\item Representation Learning: this work used \texttt{graph2vec} as the main procedure to learn the distributed representations. The choice was supported by its theoretically applicability to this work and motivated by existing related works on CF graph metafeatures~\citep{Cunha2018graphs}. However, other techniques can be considered, such as Autoencoders (for instance, adapting Image Processing techniques using Convolutional Neural Networks~\citep{NIPS1989_293} to the CF domain, given the similarity between images and rating matrices) and RL techniques specially designed for CF~\citep{Sedhain2015,DBLP:journals/corr/StrubMG16a,Wu2016} (but designed to describe the whole dataset rather than each user; alternatively, these representations can be directly used to perform algorithm selection on an user level). Despite the clear motivation to use either, none is yet ready to be applied to the general CF algorithm selection problem and require modifications.
\item Hyperparameter Tuning: according to the experimental results, the proposed technique is sensitive to the value of some hyperparameters. Hence, the metafeature extraction process requires training multiple \texttt{graph2vec} models to find the best one. Although this study tried to indicate the best hyperparameter values (which may be used as default), it is essential to understand that a different experimental setup may require a different hyperparameter setting to achieve optimal results.
\item Predictive Performance: the presented experimental setup, although extensive in nature, may still be the reason to why the proposed metafeatures are not significantly better than the state-of-the-art. Reasons such as insufficient amount of datasets and baselearners, imbalanced data (not enough meta-examples for all metatargets considered) and lack of hyperparameter optimisation in the baselearners may influence the experimental results. However, the authors would like to acknowledge how difficult it still is to obtain a more suitable experimental setup in the CF domain. 
\item Metafeature Importance: most previous works in CF algorithm selection have investigated which metafeatures are the most relevant. Despite becoming increasingly complex and harder to interpret, there was still a hint towards which data properties were important. In this case, since the metafeatures created are latent, it is impossible to perform the same analysis. Therefore, until procedures able to extract meaning from latent features surface, the analysis is limited to the one presented in this work.
\end{itemize}

\bibliographystyle{spbasic}      
\bibliography{myrefs}   

\begin{thebibliography}{63}
\providecommand{\natexlab}[1]{#1}
\providecommand{\url}[1]{{#1}}
\providecommand{\urlprefix}{URL }
\expandafter\ifx\csname urlstyle\endcsname\relax
  \providecommand{\doi}[1]{DOI~\discretionary{}{}{}#1}\else
  \providecommand{\doi}{DOI~\discretionary{}{}{}\begingroup
  \urlstyle{rm}\Url}\fi
\providecommand{\eprint}[2][]{\url{#2}}

\bibitem[{Adomavicius and Zhang(2012)}]{Adomavicius2012}
Adomavicius G, Zhang J (2012) {Impact of data characteristics on recommender
  systems performance}. ACM Trans Manag Inf Syst 3(1):1--17

\bibitem[{Amasyali and Ersoy(2009)}]{Amasyali2009}
Amasyali F, Ersoy O (2009) {A study of meta learning for regression}. Tech.
  rep., Purdue University

\bibitem[{Bengio(2011)}]{Bengio2011}
Bengio Y (2011) {Deep Learning of Representations for Unsupervised and Transfer
  Learning}. JMLR: Workshop and Conference 7:1--20

\bibitem[{Bengio et~al(2013)Bengio, Courville, and Vincent}]{Bengio2012}
Bengio Y, Courville A, Vincent P (2013) {Representation Learning: Review and
  New Perspectives}. IEEE Trans Pattern Anal Mach Intell 35(8):1798--1828

\bibitem[{Bergstra and Bengio(2012)}]{BergstraJAMESBERGSTRA2012}
Bergstra J, Bengio Y (2012) {Random Search for Hyper-Parameter Optimization}.
  JMLR 13:281--305

\bibitem[{Bourlard and Kamp(1988)}]{Bourlard1988}
Bourlard H, Kamp Y (1988) {Auto-association by multilayer perceptrons and
  singular value decomposition}. Biological Cybernetics 59(4):291--294

\bibitem[{Brazdil and Soares(2000)}]{Brazdil2000}
Brazdil P, Soares C (2000) {A Comparison of Ranking Methods for Classification
  Algorithm Selection}. In: de~M{\'{a}}ntaras R, Plaza E (eds) Machine
  Learning: ECML 2000, Springer Berlin Heidelberg, pp 63--75

\bibitem[{Brazdil et~al(2003)Brazdil, Soares, and Costa}]{Brazdil2003}
Brazdil P, Soares C, Costa J (2003) {Ranking Learning Algorithms: Using IBL and
  Meta-Learning on Accuracy and Time}. Machine Learning 50(3):251--277

\bibitem[{Brazdil et~al(2009)Brazdil, Giraud-Carrier, Soares, and
  Vilalta}]{Brazdil2009}
Brazdil P, Giraud-Carrier C, Soares C, Vilalta R (2009) {Metalearning:
  Applications to Data Mining}, 1st edn. Springer Publishing

\bibitem[{Collins et~al(2018)Collins, Beel, and Tkaczyk}]{Collins2018}
Collins A, Beel J, Tkaczyk D (2018) {One-at-a-time: A Meta-Learning
  Recommender-System for Recommendation}. ArXiv e-prints
  \eprint{arXiv:1805.12118}

\bibitem[{Cunha et~al(2016)Cunha, Soares, and Carvalho}]{Cunha2016}
Cunha T, Soares C, Carvalho A (2016) {Selecting Collaborative Filtering
  algorithms using Metalearning}. In: ECML-PKDD, pp 393--409

\bibitem[{Cunha et~al(2017{\natexlab{a}})Cunha, Soares, and
  Carvalho}]{Cunha:2017:MCF:3109859.3109899}
Cunha T, Soares C, Carvalho A (2017{\natexlab{a}}) {Metalearning for
  Context-aware Filtering: Selection of Tensor Factorization Alg}. In: ACM
  RecSys, pp 14--22

\bibitem[{Cunha et~al(2017{\natexlab{b}})Cunha, Soares, and
  Carvalho}]{Cunha2017}
Cunha T, Soares C, Carvalho A (2017{\natexlab{b}}) {Recommending Collaborative
  Filtering algorithms using landmarkers}. In: Discovery Science, pp 189--203

\bibitem[{Cunha et~al(2018{\natexlab{a}})Cunha, Soares, and
  Carvalho}]{Cunha2018graphs}
Cunha T, Soares C, Carvalho A (2018{\natexlab{a}}) {Algorithm Selection for
  Collaborative Filtering: the influence of graph metafeatures and
  multicriteria metatargets}. ArXiv e-prints pp 1--25,
  \eprint{arXiv:1807.09097}

\bibitem[{Cunha et~al(2018{\natexlab{b}})Cunha, Soares, and
  Carvalho}]{Cunha2018128}
Cunha T, Soares C, Carvalho A (2018{\natexlab{b}}) {Metalearning and
  Recommender Systems: A literature review and empirical study on the algorithm
  selection problem for Collaborative Filtering}. Information Sciences
  423:128--144

\bibitem[{Dem{\v{s}}ar(2006)}]{Demsar2006}
Dem{\v{s}}ar J (2006) {Statistical Comparisons of Classifiers over Multiple
  Data Sets}. Journal of Machine Learning Research 7:1--30

\bibitem[{Dooms et~al(2013)Dooms, {De Pessemier}, and
  Martens}]{Dooms13crowdrec}
Dooms S, {De Pessemier} T, Martens L (2013) {MovieTweetings: a Movie Rating
  Dataset Collected From Twitter}. In: CrowdRec at ACM RecSys

\bibitem[{Ekstrand and Riedl(2012)}]{Ekstrand2012}
Ekstrand M, Riedl J (2012) {When Recommenders Fail: Predicting Recommender
  Failure for Algorithm Selection}. ACM RecSys pp 233--236

\bibitem[{Gama and Brazdil(1995)}]{Gama1995}
Gama J, Brazdil P (1995) {Characterization of Classification Algorithms}.
  Lecture Notes in Computer Science 990:189--200

\bibitem[{Gantner et~al(2011)Gantner, Rendle, Freudenthaler, and
  Schmidt-Thieme}]{Gantner2011}
Gantner Z, Rendle S, Freudenthaler C, Schmidt-Thieme L (2011) {MyMediaLite: A
  Free Recommender System Library}. In: ACM RecSys, pp 305--308

\bibitem[{Goldberg et~al(2001)Goldberg, Roeder, Gupta, and
  Perkins}]{Goldberg2001}
Goldberg K, Roeder T, Gupta D, Perkins C (2001) {Eigentaste: A Constant Time
  Collaborative Filtering Algorithm}. Information Retrieval 4(2):133--151

\bibitem[{Goodfellow et~al(2016)Goodfellow, Bengio, and
  Courville}]{Goodfellow-et-al-2016}
Goodfellow I, Bengio Y, Courville A (2016) Deep Learning. MIT Press

\bibitem[{Griffith et~al(2012)Griffith, O'Riordan, and Sorensen}]{Griffith2012}
Griffith J, O'Riordan C, Sorensen H (2012) {Investigations into rating
  information and accuracy in collaborative filtering}. In: ACM SAC, pp
  937--942

\bibitem[{GroupLens(2016)}]{GroupLens2016}
GroupLens (2016) {MovieLens}.
  \urlprefix\url{http://grouplens.org/datasets/movielens/}

\bibitem[{He et~al(2016)He, Zhang, Ren, and Sun}]{He2016770}
He K, Zhang X, Ren S, Sun J (2016) {Deep residual learning for image
  recognition}. In: IEEE Comput Soc Conf Comput Vis Pattern Recognit, pp
  770--778

\bibitem[{Hu et~al(2008)Hu, Koren, and Volinsky}]{Hu2008a}
Hu Y, Koren Y, Volinsky C (2008) {Collaborative Filtering for Implicit Feedback
  Datasets}. In: IEEE Int. Conf. on Data Mining, pp 263 -- 272

\bibitem[{H{\"{u}}llermeier et~al(2008)H{\"{u}}llermeier, F{\"{u}}rnkranz,
  Cheng, and Brinker}]{Hullermeier2008}
H{\"{u}}llermeier E, F{\"{u}}rnkranz J, Cheng W, Brinker K (2008) {Label
  ranking by learning pairwise preferences}. Artificial Intelligence
  172(16-17):1897--1916

\bibitem[{Kalousis and Hilario(2001)}]{10.1007/3-540-45357-1_26}
Kalousis A, Hilario M (2001) {Feature Selection for Meta-learning}. In:
  Advances in Knowledge Discovery and Data Mining, pp 222--233

\bibitem[{Koren(2008)}]{Koren2008}
Koren Y (2008) {Factorization meets the neighborhood: a multifaceted
  collaborative filtering model}. In: ACM SIGKDD, pp 426--434, \eprint{62}

\bibitem[{Koren(2010)}]{Koren2010}
Koren Y (2010) {Factor in the Neighbors: Scalable and Accurate Collaborative
  Filtering}. ACM Trans Knowl Discov Data 4(1):1--24

\bibitem[{Le and Mikolov(2014)}]{Le:2014:DRS:3044805.3045025}
Le Q, Mikolov T (2014) {Distributed Representations of Sentences and
  Documents}. In: Int. Conf. on Machine Learning, pp II--1188----II--1196

\bibitem[{Lecun(1987)}]{37f2b6bee745402aa4e4d124d33be0e0}
Lecun Y (1987) {PhD thesis: Modeles connexionnistes de l'apprentissage
  (connectionist learning models)}. Universite P. et M. Curie (Paris 6)

\bibitem[{LeCun et~al(1990)LeCun, Boser, Denker, Henderson, Howard, Hubbard,
  and Jackel}]{NIPS1989_293}
LeCun Y, Boser B, Denker J, Henderson D, Howard R, Hubbard W, Jackel L (1990)
  {Handwritten Digit Recognition with Back-Propagation Network}. In: Adv. in
  Neural Information Processing Sys., Morgan Kaufmann, pp 396--404

\bibitem[{Lecun et~al(2015)Lecun, Bengio, and Hinton}]{Lecun2015}
Lecun Y, Bengio Y, Hinton G (2015) {Deep learning}. Nature 521(7553):436--444

\bibitem[{Leskovec and Faloutsos(2006)}]{Leskovec2006}
Leskovec J, Faloutsos C (2006) Sampling from large graphs. In: ACM SIGKDD, pp
  631--636

\bibitem[{Matuszyk and Spiliopoulou(2014)}]{Matuszyk2014}
Matuszyk P, Spiliopoulou M (2014) {Predicting the Performance of Collaborative
  Filtering}. In: Web Intelligence, Mining and Semantics, pp 38:1--6

\bibitem[{McAuley and Leskovec(2013)}]{McAuley2013}
McAuley J, Leskovec J (2013) {Hidden Factors and Hidden Topics: Understanding
  Rating Dimensions with Review Text}. In: ACM RecSys, pp 165--172

\bibitem[{Menon and Elkan(2010)}]{Menon2010}
Menon AK, Elkan C (2010) {A log-linear model with latent features for dyadic
  prediction}. In: ICDM, pp 364--373

\bibitem[{Mikolov et~al(2013)Mikolov, Chen, Corrado, and Dean}]{Mikolov2013}
Mikolov T, Chen K, Corrado G, Dean J (2013) {Efficient Estimation of Word
  Representations in Vector Space}. ArXiv e-prints pp 1--12,
  \eprint{arXiv:1301.3781}

\bibitem[{Narayanan et~al(2017)Narayanan, {Chandramohan, Mahinthan Venkatesan},
  Chen, Liu, and Jaiswal}]{Narayanan2017}
Narayanan A, {Chandramohan, Mahinthan Venkatesan} R, Chen L, Liu Y, Jaiswal S
  (2017) {graph2vec: Learning Distributed Representations of Graphs}. ArXiv
  e-prints pp 1--8, \eprint{arXiv:1707.05005}

\bibitem[{Paterek(2007)}]{Paterek2007}
Paterek A (2007) {Improving regularized singular value decomposition for
  collaborative filtering}. In: KDD cup and workshop, pp 2--5

\bibitem[{Pinto et~al(2016)Pinto, Soares, and Mendes-Moreira}]{Pinto2016}
Pinto F, Soares C, Mendes-Moreira J (2016) {Towards automatic generation of
  Metafeatures}. In: PAKDD, pp 215--226

\bibitem[{Prud{\^{e}}ncio and Ludermir(2004)}]{Prudencio2004}
Prud{\^{e}}ncio RBC, Ludermir TB (2004) {Meta-learning approaches to selecting
  time series models}. Neurocomputing 61:121--137

\bibitem[{Rendle et~al(2009)Rendle, Freudenthaler, Gantner, and
  Schmidt-thieme}]{Rendle2009}
Rendle S, Freudenthaler C, Gantner Z, Schmidt-thieme L (2009) {BPR: Bayesian
  Personalized Ranking from Implicit Feedback}. In: Conference on Uncertainty
  in Artificial Intelligence, pp 452--461

\bibitem[{Rice(1976)}]{DBLP:journals/ac/Rice76}
Rice J (1976) {The Algorithm Selection Problem}. Adv in Computers 15:65--118

\bibitem[{Rumelhart et~al(1986)Rumelhart, McClelland, and PDP
  Research~Group}]{Rumelhart:1986:PDP:104279}
Rumelhart DE, McClelland JL, PDP Research~Group C (eds)  (1986) Parallel
  Distributed Processing: Explorations in the Microstructure of Cognition, Vol.
  1: Foundations. MIT Press, Cambridge, MA, USA

\bibitem[{Salakhutdinov and Mnih(2008)}]{Salakhutdinov2008}
Salakhutdinov R, Mnih A (2008) {Probabilistic Matrix Factorization.} In:
  Advances in Neural Information Processing Systems, pp 1257--1264

\bibitem[{Schmidhuber(2015)}]{Schmidhuber2015}
Schmidhuber J (2015) {Deep learning in neural networks: An overview}. Neural
  Networks 61:85--117

\bibitem[{Sedhain et~al(2015)Sedhain, Menon, Sanner, and Xie}]{Sedhain2015}
Sedhain S, Menon AK, Sanner S, Xie L (2015) {AutoRec : Autoencoders Meet
  Collaborative Filtering}. In: WWW, pp 111--112

\bibitem[{Shervashidze et~al(2011)Shervashidze, Schweitzer, van Leeuwen,
  Mehlhorn, and Borgwardt}]{Shervashidze:2011:WGK:1953048.2078187}
Shervashidze N, Schweitzer P, van Leeuwen EJ, Mehlhorn K, Borgwardt KM (2011)
  {Weisfeiler-Lehman Graph Kernels}. JMLR 12:2539--2561

\bibitem[{Smith{-}Miles(2008)}]{Smith-Miles08}
Smith{-}Miles K (2008) Cross-disciplinary perspectives on meta-learning for
  algorithm selection. {ACM} Comput Surv 41(1):6:1--6:25

\bibitem[{Soares(2015)}]{Soares2015}
Soares C (2015) {labelrank: Predicting Rankings of Labels}.
  \urlprefix\url{https://cran.r-project.org/package=labelrank}

\bibitem[{Strub et~al(2016)Strub, Mary, and
  Gaudel}]{DBLP:journals/corr/StrubMG16a}
Strub F, Mary J, Gaudel R (2016) {Hybrid Recommender System based on
  Autoencoders}. ArXiv e-prints \eprint{arxiv:1606.07659}

\bibitem[{{Van Der Maaten} and Hinton(2008)}]{VanDerMaaten2008}
{Van Der Maaten} LJP, Hinton GE (2008) {Visualizing high-dimensional data using
  t-sne}. Journal of Machine Learning Research 9:2579--2605

\bibitem[{Vanschoren(2010)}]{Vanschoren2010}
Vanschoren J (2010) {Understanding machine learning performance with experiment
  databases}. PhD thesis, Katholieke Universiteit Leuven

\bibitem[{Vembu and G{\"{a}}rtner(2010)}]{Vembu2010}
Vembu S, G{\"{a}}rtner T (2010) {Label ranking algorithms: A survey}. In:
  Preference Learning, SpringerVerlag, pp 45--64

\bibitem[{Wang et~al(2011)Wang, Lu, and Zhai}]{Wang2011}
Wang H, Lu Y, Zhai C (2011) {Latent Aspect Rating Analysis Without Aspect
  Keyword Supervision}. In: ACM SIGKDD, pp 618--626

\bibitem[{Weimer et~al(2008)Weimer, Karatzoglou, and Smola}]{Weimer2008}
Weimer M, Karatzoglou A, Smola A (2008) {Improving Maximum Margin Matrix
  Factorization}. Machine Learning 72(3):263--276

\bibitem[{Wu et~al(2016)Wu, DuBois, Zheng, and Ester}]{Wu2016}
Wu Y, DuBois C, Zheng AX, Ester M (2016) {Collaborative Denoising Auto-Encoders
  for Top-N Recommender Systems}. In: WSDM, pp 153--162

\bibitem[{Yahoo!(2016)}]{Yahoo}
Yahoo! (2016) {Webscope datasets}.
  \urlprefix\url{https://webscope.sandbox.yahoo.com/}

\bibitem[{Yelp(2016)}]{Yelp2016}
Yelp (2016) {Yelp Dataset}.
  \urlprefix\url{https://www.yelp.com/dataset\_challenge}

\bibitem[{Zafarani and Liu(2009)}]{Zafarani+Liu:2009}
Zafarani R, Liu H (2009) Social computing data repository at {ASU}.
  \urlprefix\url{http://socialcomputing.asu.edu}

\bibitem[{Ziegler et~al(2005)Ziegler, McNee, Konstan, and Lausen}]{Ziegler2005}
Ziegler CN, McNee SM, Konstan JA, Lausen G (2005) {Improving Recommendation
  Lists Through Topic Diversification}. In: WWW, pp 22--32

\end{thebibliography}

\end{document}